\documentclass[]{spie}  
\usepackage{url}

\usepackage{amsmath,amsfonts,amssymb}
\usepackage{graphicx}

\usepackage[colorlinks=true, allcolors=blue]{hyperref}

\title{Fabrication of focusing optics for TA-MOONS: Micro-MOONS}

\author[a]{Ceiwynn Longworth} 
\author[a,b]{Megan Delamer}
\author[a,b]{Suvrath Mahadevan}
\author[c]{Joe P. Ninan}
\author[d]{Kathleen Gehoski}
\author[c]{Krushna Jadhav} 

\affil[a]{Department of Astronomy \& Astrophysics, Penn State, University Park, State College, USA}
\affil[b]{Center for Exoplanets \& Habitable Worlds,  Penn State, University Park, State College, USA}
\affil[c]{Department of Astronomy and Astrophysics, Tata Institute of Fundamental Research, Homi Bhabha Road, Colaba, Mumbai 400005, India}
\affil[d]{Materials Research Institute, Penn State, University Park, State College, USA}

\authorinfo{Further author information: (Send correspondence to C.L.)\\C.L.: E-mail: ccl5356@psu.edu}

\usepackage{hyperref}
\begin{document} 
\maketitle

\begin{abstract}

~~~~Understanding star and planet formation requires optical to near-infrared spectroscopic observations of a large number of young stellar objects. The TIFR-ARIES Multi-Object Optical to Near-infrared Spectrometer’s (TA-MOONS) primary objective is to perform a large spectroscopic survey of young stellar objects across wavelengths of 360 nm to 2.5 $\mu$m, with the multiplexed capability of observing up to eight sources simultaneously. Multiplexity is achieved by moving pickup mirrors on robotic arms. One component of this robotic arm system is Micro-MOONS, a micro-optic array, nanofabricated to enable accurate calibration, validation, and positioning of stellar images onto the pickup mirrors. The micro-mirrors of the Micro-MOONS system are spherically concave, fabricated via two-photon polymerization (2PP) in the NanoScribe GT2 system, and are printed on plasma-cleaned silicon substrates using IP-S photoresist, followed by gold coating. 2PP allows for fine, detailed shapes on a sub-$\mu m$  scale that are otherwise difficult and costly to obtain. The optimization of Micro-MOONS focuses on achieving uniformity, minimizing surface roughness, and preserving the desired radius of curvature to avoid beam obstruction near the pickup mirrors.  This work demonstrates the feasibility of integrating advanced custom-printed micro-optics into multi-object spectroscopic instruments, enabling improved calibration and efficiency in large-scale stellar surveys.
\end{abstract}


\section{INTRODUCTION}

~~~~Protoplanetary disks are the birthplaces of all planets, but persist for only a few million years\cite{Polnitzky_precise_2026}, leaving open questions about how the gas and dust within the disk can coalesce into planets on such rapid timescales. While those timescales are short in an astrophysical sense, a singular system cannot be observed across any meaningful fraction of its lifetime; each one instead provides a snapshot at a singular time. Much of the work of reconciling the differences between the planet population expected to arise from observed disks and the actual observed planet population falls to modeling and simulation\cite{Emsenhuber_ngpps_2021,Burn_ngpps_2021}. Key inputs in accurately modeling the process are the thermal and chemical environments of the disk, both of which are impacted by the host star. While observations with the Atacama Large Millimeter/sub-millimeter Array (ALMA) and other radio telescopes can provide this information for the cold outer disks where giant planets most likely form\cite{Guzman_maps_2021,Martire_rotation_2024} --- including direct observations of forming giant planets\cite{Keppler_highly_2019}, accessing the inner portions of the disk where Earth-like terrestrial planets formed requires shorter wavelengths. Observations of inner disks with the James Webb Space Telescope (JWST) at mid-IR wavelengths have revealed rich chemistry\cite{Colmenares_jwst/miri_2024,Grant_minds_2024}, but expanding the observations bluewards gives access to atomic lines and molecular bandheads that trace accretion and stellar outflows. These lines can provide information on the kinematic and opacity signatures in the disk environment, as well as the underlying stellar spectrum that influences distributions of molecular species within the disk. Energetic phenomena like FU Orionis outbursts still remain poorly understood, but can influence the physical properties and chemistry of these disks.

The Tata Institute of Fundamental Research (TIFR) and  Aryabhatta Research Institute of Observational Sciences (ARIES) are leading the development of a multi-object spectrograph for the 3.6\,m Devasthal Optical Telescope (DOT)\cite{Sagar_DOT_2019} to conduct a large survey of the active inner regions of disks around young stellar objects (YSOs). The TIFR-ARIES Multi-Object Optical to Near Infrared Spectrometer (TA-MOONS) will have continuous wavelength coverage from the Balmer break at 360 nm to CO lines at 2.5$\mu m$ with a resolution of R$\sim$2700, which excludes silica optical fiber-based architectures due to attenuation at the bluest and reddest wavelengths \cite{Shah_Mechanical_2024}. TA-MOONS instead uses a novel deployable slit (DS) technique, wherein a series of mirror-based arms with translational and rotational motion can each pick up a single source within an annular subset of the 12 arcmin field of view (FOV) \cite{Poojary_Front-Optics_2024}. The mechanical design of the individual arms is similar to that in the K-band Multi-Object Spectrograph (KMOS)\cite{Sharples_KMOST_2004,Bennett_mechanical_2008} on the Very Large Telescope (VLT). Each arm of TA-MOONS will have an independent cryo shutter, enabling simultaneous observations of objects with varying brightnesses, and the 6" long deployable slit can also be used for extended source observation along arbitrary orientations \cite{Shah_Mechanical_2024}.

The design requires 10 $\mu$m accuracy in the placement of the pick-up arm on the star, which can be achieved by simultaneously imaging the probes and the star field with a single 4kx4k CMOS array. The 10 $\mu$m accuracy requirements is a focal-plane mechanical tolerance derived from the inner-working mechanics of the instrumentation so as to avoid significant loss of flux. The tolerance is calibrated to the theoretical worst credible seeing condition that the telescope would routinely operate in, allowing enough margin of error that the mechanics never become the dominant source of flux loss. At the DOT's 32.4 m effective focal length, and plate scale of $\sim$6.4 arcsec/mm, the 10$\mu$m corresponds to $\sim$0.064 arcseconds.

\begin{figure}
    \centering
    \includegraphics[width=0.95\linewidth]{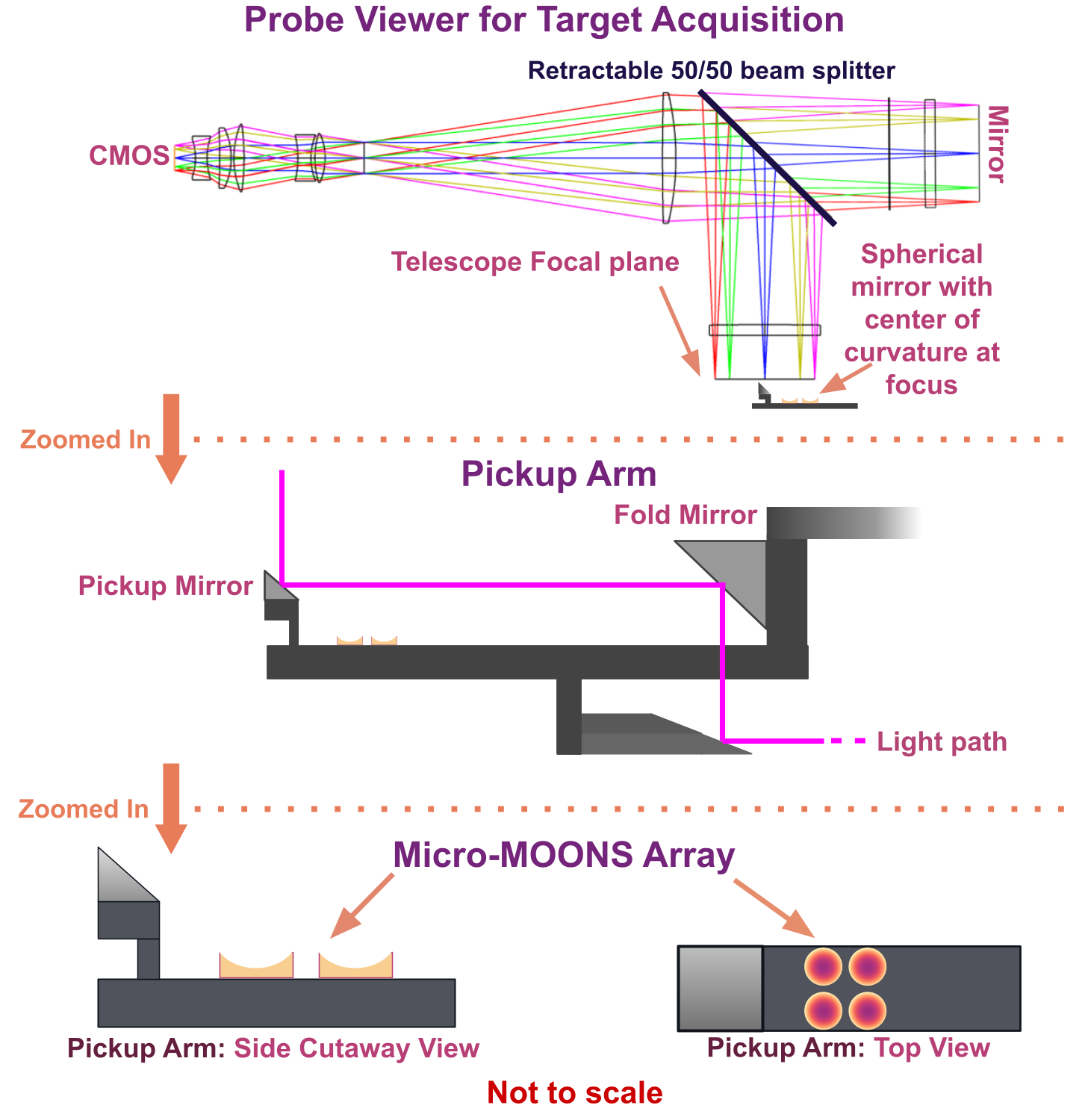}
    \caption{ Is a diagram of how the Micro-MOONS array works within the TA-MOONS’s probe viewer system. Showing also a model of a single pickup arm from the TA-MOONS instrument design, demonstrating how light reflects through the system. Circled at the end of the arm is where the micro-MOONS project will be located. Finally a model of the approximate position and organization of the micro-MOONS arrays on a pickup arm}
    \label{fig:TA-MOONS & Micro-MOONS Mechanical Overview}
\end{figure}

\begin{figure}
    \centering
    \includegraphics[width=0.85\linewidth]{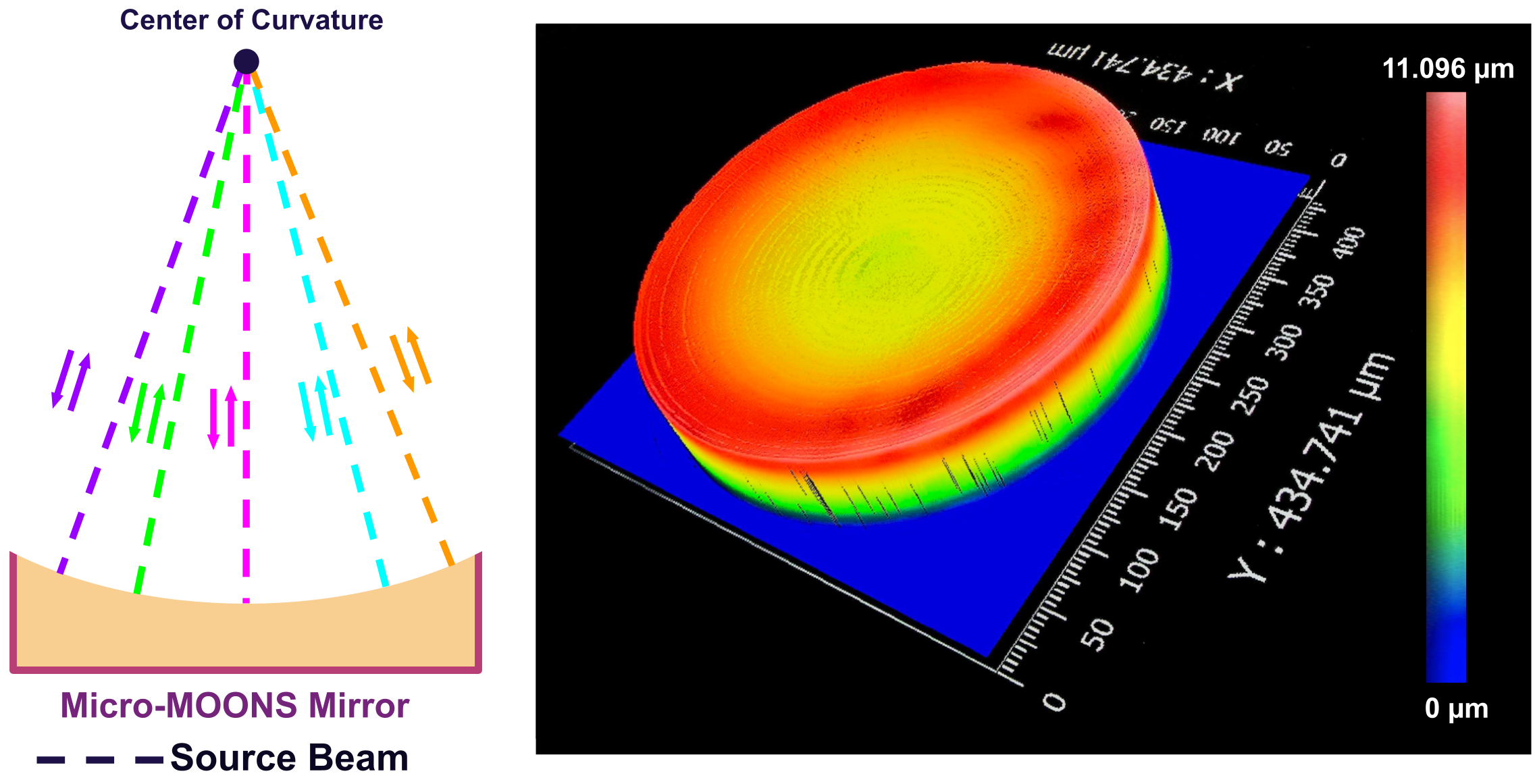}
    \caption{The left image demonstrates how reflection occurs with the Micro-MOONS mirror, and the right image is of a 3D model of a micro mirror imaged using an Optical Profilometer in the Penn State Millennium Science Complex.}
    \label{fig:Micro-MOONS Mirror Reflection Overview}
\end{figure}

The accuracy of the placement is confirmed by precisely measuring the stellar position along a fiducial point on the arm using a small array of concave spherical mirrors with centers of curvature on the telescope focal plane (see \autoref{fig:TA-MOONS & Micro-MOONS Mechanical Overview}). These mirrors act as position-sensitive retro-reflectors; a star imaged on any one of the mirrors will appear as a bright spot in the CMOS array, providing absolute position calibration of the pick-up arm. By making the mirrors spherical, we render them insensitive to the tip/tilt of the pick up arm (see \autoref{fig:Micro-MOONS Mirror Reflection Overview}).

Two photon polymerization (2PP)\cite{goppert_uber_1931,maruo_three_1997} based 3D printing offers the flexibility to create bespoke micro-optics for both transmission and reflection at scales or with geometries that standard fabrication techniques would struggle to replicate. In this technique, a femtosecond laser is focused into a pool of liquid photoresist via a high numerical aperture objective. The rapid pulses allow for two different photons to interact with the same photoactivated initiator nearly simultaneously: the first excites an electron to a short-lived virtual intermediate state, and the second kicks the electron off, allowing for the creation of a free radical and beginning the polymerization process (see O'Halloran\cite{OHalloran_2pp_2023} and the references therein). While a single, higher energy photon can also induce polymerization, the reaction can occur anywhere along the path of the laser; the cross section in two photon absorption produces a very small print region (or voxel) where there is sufficient energy to trigger polymerization\cite{fischer_three_2013}. The needed surface tolerance on the printed mirrors based on seeing at DOT is $\sim\lambda/4$ in the I band, which is well within the capabilities of 2PP. Given the large number of mirrors needed (4 mirrors in each of the eight deployable slit arms), and the small size and radius of the mirrors needed so as not to interfere with pick off optics, 2PP is a flexible and viable solution for this compared to conventional optics manufacturing. In this work, we present a summary of our work at Penn State in fabricating, coating, and characterizing the spherical mirror arrays in preparation for their eventual inclusion in TA-MOONS.

\begin{figure}
    \centering
    \includegraphics[width=0.95\linewidth]{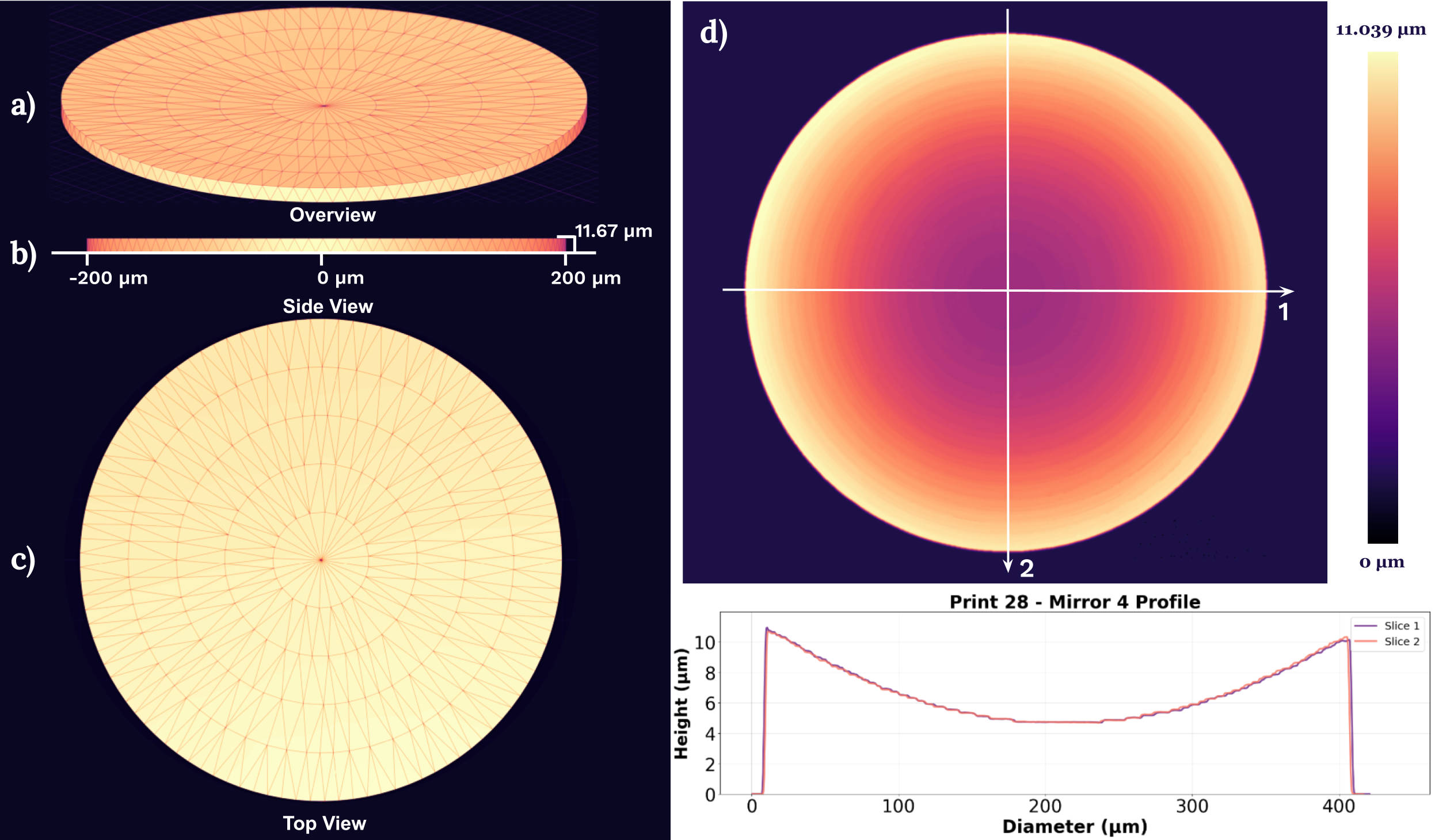}
    \caption{ Overview of the CAD model and print of a micro-mirror. a) An overview of the CAD design. b) Side-profile view, including a scale to demonstrate size in $\mu m$ s. c) A top-down view of the CAD design of the concave mirror. All CAD images were generated with Autodesk Fusion. d) Shows an intensity image of a printed micro-mirror, top-down topography data. The graph below is a profile plot taking data from the slices through the surface intensity graph above.}
    \label{fig:Overview of a Micro-MOONS mirror design}
\end{figure}

\section{METHODS}
\subsection{Design}
~~~~Micro-MOONS is a 2x2 array of spherical, concave mirrors that are 400 $\mu m$ in diameter with a 500 $\mu m$  pitch from one mirror center to the next (See \autoref{fig:Overview of a Micro-MOONS mirror design}). The acceptable range of radius of curvature is 3-4 mm, with the requirement that the radii mirrors in each set be within $\pm 100\,\mu m$  of each other. While the sets on different arms could be placed at slightly different heights with shims, the four mirrors in each arm could not be individually adjusted. The starting design height and radius of curvature (ROC) are 11.67 $\mu m$ and 3000 $\mu m$, respectively.  The height contains a sag of 6.67 $\mu m$ and a base height of 5 $\mu m$. The mirrors are solid prints using 2PP, with the photoresist IP-S on a silicon substrate. We replicate this array multiple times to include 8 final arrays, one for each pick-up arm of TA-MOONS, and some spares. All arrays are coated with a layer of gold to maximize reflexivity into the infrared, given that YSOs, the main targets, are heavily reddened. Gold also provides significantly higher reflectivity than aluminum in the astronomical I band, which the guiding optics of TIFR-MOONS are optimized for, and where aluminum exhibits a reflectivity dip.

\subsection{Fabrication Overview}
~~~Our fabrication overview is as follows: Prints begin with a silicon substrate that is chemically washed, blown dry with nitrogen gas, and then plasma etched. After cleaning the substrate of undesired particles and residue, we dispense IP-S resin onto the surface. Once dispensed, the resist and substrate thermalize, followed by a 30 minute de-gassing period with intervals of burping. Before loading the print tray into the NanoScribe Photonic Professional GT 3D printer (NanoScribe). After we unload and heat-cure the substrate then develop it in a chemical bath of mr-Dev 600, followed by a 2-minute IPA soak. Micro-mirror fabrication is concluded with a UV curing period.

\subsubsection{Fabrication}
~~~~The micro-MOONS mirrors are first created as a Computer-Aided Design (CAD) file using Autodesk Fusion. Sag is dependent on our desired ROC and diameter, while the 5~$\mu m$   base was chosen to sufficiently buffer the sag from the substrate, as well as further promote adhesion. The CAD file is exported as a solid component .stl and uploaded to the proprietary software DeScribe, which is utilized by the Photonic Professional printer, to convert the file into generalized writing language (.gwl) files which are readable by the printer. Within the DeScribe (we used version 2.6) software, we can edit the size, parameters, and array style, optimizing the print job to our specifications.

The NanoScribe Photonic Professional GT is a 3D printer that utilizes 2PP to print microscale designs, located within the nanofabrication facility in the Materials Research Institute at the Pennsylvania State University.  It uses a 780 nm femtosecond laser with an average output power of 180 mW, and a pulse duration of 80-100 fs\cite{bunea_Micro_2021}. There are 3 different objectives to choose from to print, but for our purposes, we use the 25x objective as it's capable of printing small features and fitting one micro mirror into it's working distance of 400 $\mu m$ s, able to print a solid 500 $\mu m$  diameter spherically concave mirror\footnote{https://www.epfl.ch/research/facilities/cmi/equipment/photolithography/nanoscribe-photonic-professional-gt/}. By removing the need to move the stage during a singular mirror's print, we maintain uniformity of the shape.

When converting our .stl CAD design into a printer-compatible .gwl file, we maintain the proportions of our original design, height of 11.67 $\mu m$ s, and diameter of 400 $\mu m$ s. We chose a solid printing style with a slice and hatch size of 0.2 and 0.2 $\mu m$ respectively. Slicing is the vertical distance between layers, and hatching is the distance between parallel lines within each layer. Increasing or decreasing the value of the slice and hatch affects the surface roughness of the overall design. Smaller slice and hatch distances mean more refined surfaces; larger means rougher surfaces. However, the size of the voxel sets limitations on the slice and hatch sizes. The voxel has size in two dimensions, length (l) and diameter (d), which are dependent on the laser, the objective, and the selected resin. Approximate dimensions of the voxel within our NanoScribe printer combined with the 25x objective are l= 3.3 $\mu m$ and d= 0.6 $\mu m$ \cite{delamer_pokemon_2025}. The slice and hatch of a desired design cannot exceed these sizes, or the voxel will not fully polymerize a region it passes over, as the polymerized lines will not overlap. Leaving unhardened resin that will pool out in development, risking the structural integrity of the whole design. Alternatively, a too small slice-hatch size might cause over-polymerization or potentially burning of the resist. Polymerization is dependent on dosage delivered to the photoresin, and dosage is dependent upon laser power (LP), write speed (WS), and n (an intrinsic property unique to the resin). Dosage can be modeled by the formula \cite{delamer_pokemon_2025}:  

\begin{equation}
    D=\frac{LP^n}{v}
\end{equation}

Where v is the write speed, and n is an exponent related to nonlinear absorption, ranging from 2 to 4 and dependent on the resin \cite{delamer_pokemon_2025}.   If the print is underdosed, it could be structurally unstable. Underdosing is when the aforementioned dosage to the polymer is less than the required amount to properly polymerize the resin, resulting in malformed structures. However, overdosing a section can warp the shape of the design or burn the resin. Polymerization also has a time dependency component related to the frequency of layers being printed. Rapidly printing layer after layer may cause over polymerization as a dosage is being delivered to a region in quick succession, or the alternative of under polymerization if the dosage delivered has a long wait time between layers. This time between layer printings is called the piezo settling time (PST), and is another variable that can be manipulated. Increasing PST can prevent potential over-polymerization, acting as a rest period for the printed layers.

Our substrates for printing have dimensions 25 x 25 x 0.7 mm, and are comprised of silicon, chosen for its ease of dicing, cost-effectiveness of the material, and compatibility with the IP-S photoresist. IP-S was chosen over alternative resins due to the extensive data on its behavior across all stages of the printing process, as well as its high smoothness and compatibility with the 25x objective used in our dip-in printing process. IP-S has an index of refraction of 1.479 at a wavelength of 780 nm (the NanoScribe working wavelength) and a temperature $20^{\circ}$C (the lab temperature). The NanoScribe focuses on the substrate surface (finds the interface) by detecting a refractive-index difference between the photoresist and the substrate. Silicon has an index of refraction of approximately 3.710, and IP-S has an index of refraction of 1.479, which provides a 2.221 difference. More than enough for the NanoScribe to discern between the materials, as the 25x objective needs a minimum refractive index difference of 0.1 to find the interface. 

Our main goal is to maintain a 200 $\mu m$  tolerance ($\pm$ 100 $\mu m$ range within an array) for the ROC within a single 2 x 2 array. To achieve this goal, we decided upon the parameters: 35 LP (which is 17.5 mW of power), 40k Ws, and 250 ms PST, along with a slice and hatch of 0.2 $\mu m$ with a hatching angle (HA) offset of 137.508. HA offers variability between layers, changing the direction in which printing will begin on each layer. Using this 'Golden Angle' derived from the 'Golden Ratio' as an offset angle creates non-repeating, uniform distributions. This benefits our prints by creating a more uniform final shape. We settled on the value of 137.508 so that no layer would line up exactly. The slice and hatch distance is to minimize the surface roughness of the mirrors while also minimizing print time. To increase the surface quality would have required decreasing the slice and hatch, and increasing the print time. This is non-optimal as the quality of the surface at 0.2-0.2 $\mu m$ is sufficient enough for the instrument, which requires a surface tolerance of $\sim\lambda/4$ in the I band.

Pre-processing of the sample begins with proper storage of the IP-S tube, stored upright in a refrigerated environment of approximately $4^\circ$C. The photoresist forms particulates when left in the ambient temperature of the lab, and the upright storage is meant to lower the amount of dispensed micro-bubbles. Preparation of the substrate begins with a 30-second chemical wash with acetone, followed by IPA on both sides of the silicon substrate, before blow-drying with nitrogen. We plasma etch the substrate using a SAMCO AQ 2000 for 5 minutes with $O_2$, the purpose of which is to clean the substrate and promote adhesion of the micro-mirrors. We dispense enough IP-S onto the substrate so as to allow it to form into a drop when loaded into the printer, then the substrate + the photoresist are left to sit in the Nanoscribe printer holder for 30 minutes, allowing temperatures to equalize across the substrate and resist. 
After thermalization, the substrate + printer holder goes through a 30-minute degassing stage. During which time the sample is placed into a vacuum desiccator that is attached to a Gast D0A-P135-AA compressor vacuum pump that has a maximum pressure of 60 psi. Once placed into the desiccator chamber, the lid is settled back, and the vacuum pump is turned on. Every 5 minutes over the 30 minute time-period, the valve of the desiccator is opened to re-pressurize the system; a measure of this is that the lid can be removed. This process is called 'burping' and is intended to break the bubbles that are pulled to the surface of a resist.
Once degassing is complete, the tray is removed and loaded into the NanoScribe, photoresist face down. Move the objective into the resin before the photoresist can fall onto the objective. Otherwise, there is risk of creating merger bubbles that can cause difficulty with measuring tilt and the final outcome of prints. Once we load the print tray into the NanoScribe and the 25x objective is in the resin, the interface needs to be found. Due to the highly reflective properties of silicon, focusing its surface requires an interface-minimum override: \texttt{InterfaceMin 1000.090}. Which is a code specific to the language of the printer, informing the printer to change it's requirements for when focusing on the substrate. Once the interface is found and there is a clear image of the substrate's surface on the live-view camera, measure for tilt. Due to the micro-scale nature of the print, any tilt greater than 0.05$^\circ$ is not preferred for maintaining the micro-mirrors' ROC. Prints with a tilt greater than 0.05$^\circ$ lead to errors in the profile and thus the ROC.
Once the print is done, we heat-cure the substrate for 65-95-65 $^{o}$C for 2 minutes each; the purpose of this is to further promote adhesion of the prints to the substrate. Following heat-curing, we developed the substrate in a develop bath of mr-Dev 600 for 20 minutes, which removes the unpolymerized photoresist from the substrate. Then, an IPA rinse of 2 minutes to remove any residue. Finally, we conclude with a 10-minute UV curing (duration recommended by NanoScribe for IP-S), using a COTS UV lamp\footnote{\url{https://www.melodysusie.com/products/p-plus20f-uv-protection-rechargeable-uv-led-nail-lamp}} with 48W of power and wavelengths of 365-405 nm. This finalizes the polymerization process of the printer and minimizes shrinkage of the final prints to about 10\%.

\begin{figure}
    \centering
    \includegraphics[width=0.75\linewidth]{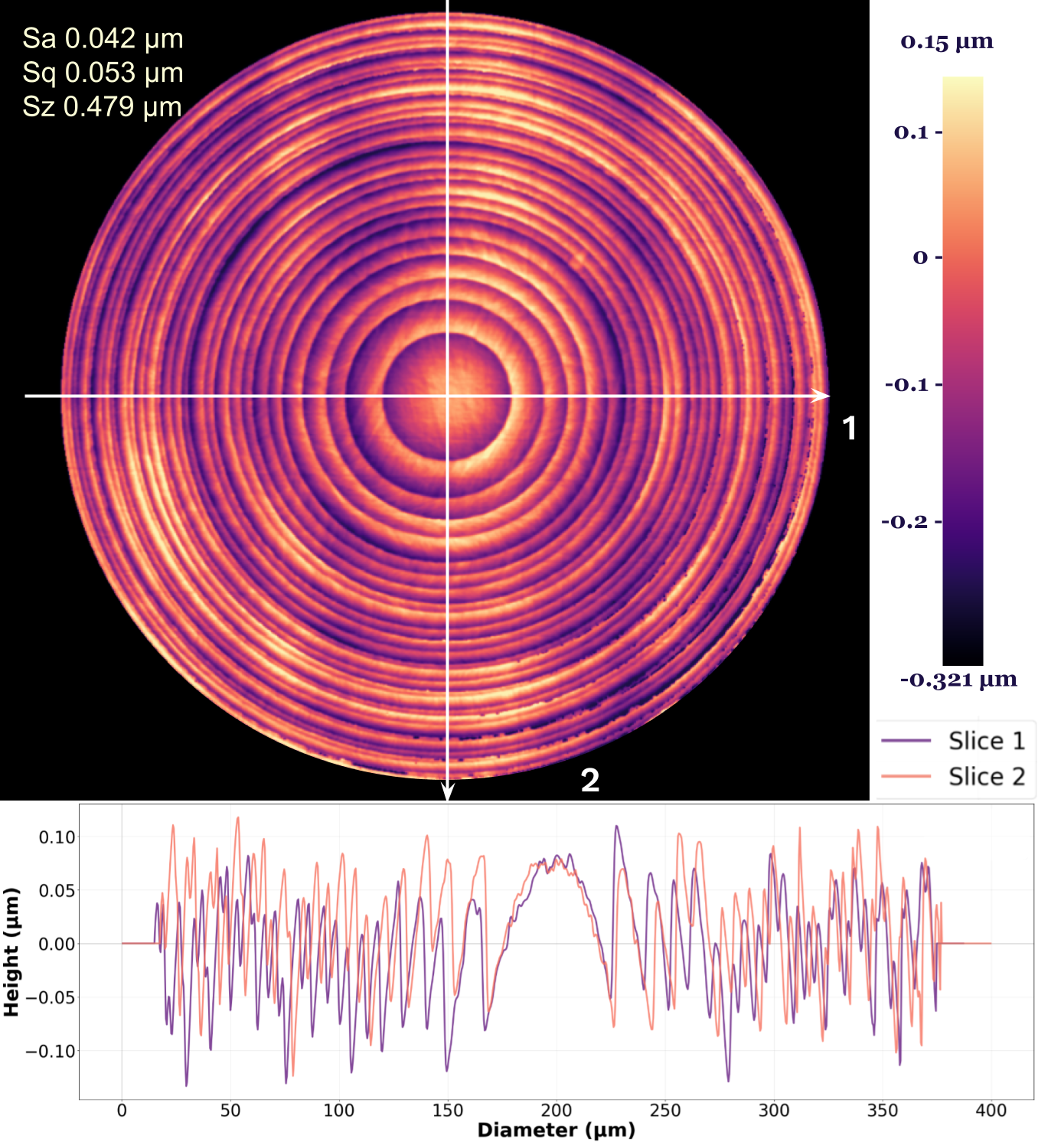}
    \caption{Intensity graph showing the inner 360-$\mu m$  diameter of a good result micro-MOONS print after subtracting a best fit circular model. The mirror has a ROC of 3212 $\mu m$ s, a maximum height deviation of 0.479 $\mu m$ from the best fit sphere. Bottom is a graph showing the slice data of the above intensity graph. A slice is a line through a region of a sample, and the subsequent graph plots the diameter and the height across the entire slice. This slice data specifically shows the mirror's surface deviation from a true sphere with a ROC of 3212 $\mu m$ s. }
    \label{fig:Spherical-deviation graph of a Singular Mirror}
\end{figure}

\begin{figure}
    \centering
    \includegraphics[width=1\linewidth]{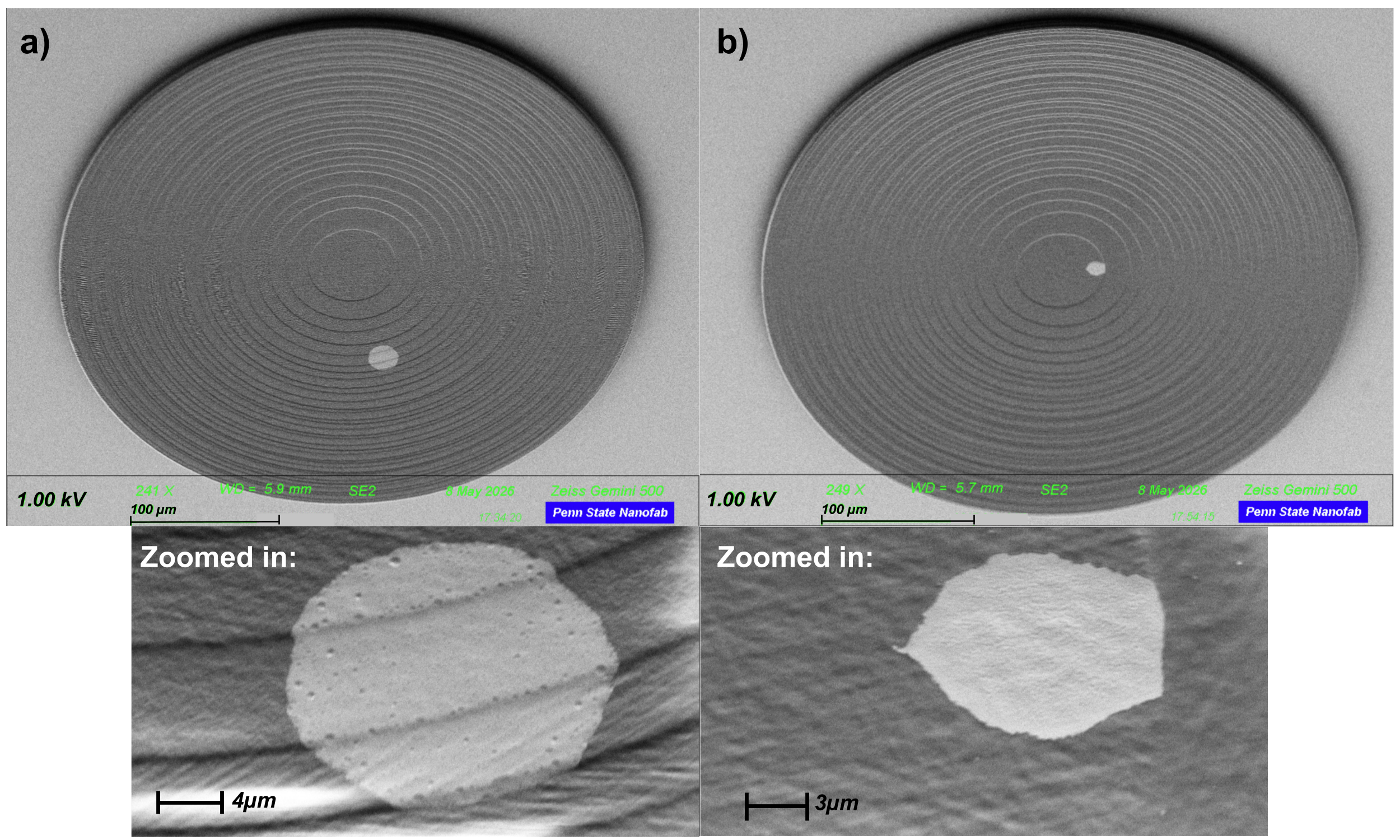}
    \caption{Images from the Field Emission Scanning Electron Microscopy: Zeiss Gemini 500 within the Penn State nanofabrication lab. The images were taken at a 30-degree tilt and 1.00 kV, with the Everhart-Thornley Detector (SE2). a) shows a mirror that was degassed for 10 minutes, while b) shows a mirror degassed for 20 minutes. The lighter gray region is a bubble that appeared during the printing process, and its size is reduced with longer degassing periods. The bubbles themselves also contain smaller micro-bubble blemishes. Their cause is still unknown to us, but we discuss our mitigation techniques.}
    \label{fig:FESEM Imgages Degassing}
\end{figure}

\subsection{Characterization}
~~~~Characterization of micro-MOONS included the use of a Nikon L200 optical microscope, a Field Emission Scanning Electron Microscopy (FESEM): Zeiss Gemini 500,  and a Zygo NexView 3D Optical Surface Profiler.
The optical microscope was used for quick post-development checks to see if there were any glaring issues, such as bubbles, burns, or poor adhesion. Optical microscope checks were quick and did not involve any measurements. 
Zygo NexView 3D Optical Surface Profiler uses white light interferometry to measure samples, and due to the reflective nature of our mirrors, there was concern regarding the imaging capabilities. However, the profilometer was fully capable of analyzing the micro-MOONS arrays. We use the 20x objective at 1x magnification, and predominantly the `Surface' setting. Occasionally, when the machine is unable to measure all data of a mirror array using the surface setting (due to errors in the print or reflection), we switch to the `surface over film' setting. The optical profilometer is used for measuring the profile, surface roughness, ROC, and deviation from spherical, using ZYGO's proprietary software, Mx. All arrays are measured the day of printing, after development, and if coated, are measured post-deposition. Arrays are measured under the optical profilometer, always before and after deposition.  \autoref{fig:Overview of a Micro-MOONS mirror design} and \autoref{fig:Spherical-deviation graph of a Singular Mirror} provide examples of data from the optical profilometer.
The FESEM was used to verify features seen under the optical profilometer that we weren't sure were real features or were caused by the machine, such as potential caverns, bubbles, or textures. The FESEM allows us to view samples from a tilt, so we can better image features. The power of the electron beam was 1.00 kV, and the aperture was changed between 15 and 20 $\mu m$ to focus on the mirrors and their smaller features. Using the FESEM, we focused on observing bubbles that had appeared in optical profilometry measurements to better understand their structure and behavior under high vacuum. As well as to confirm that they were indeed bubbles and not an imaginary feature of the profilometer. The features were indeed large bubbles containing many micro bubbles within them  and the bubbles shrunk in size after being put under high vacuum pressure in the FESEM, meaning they were unstable(see \autoref{fig:FESEM Imgages Degassing}).

\subsection{Reflective Coating Deposition}
\begin{figure}
     \centering
     \includegraphics[width=0.95\linewidth]{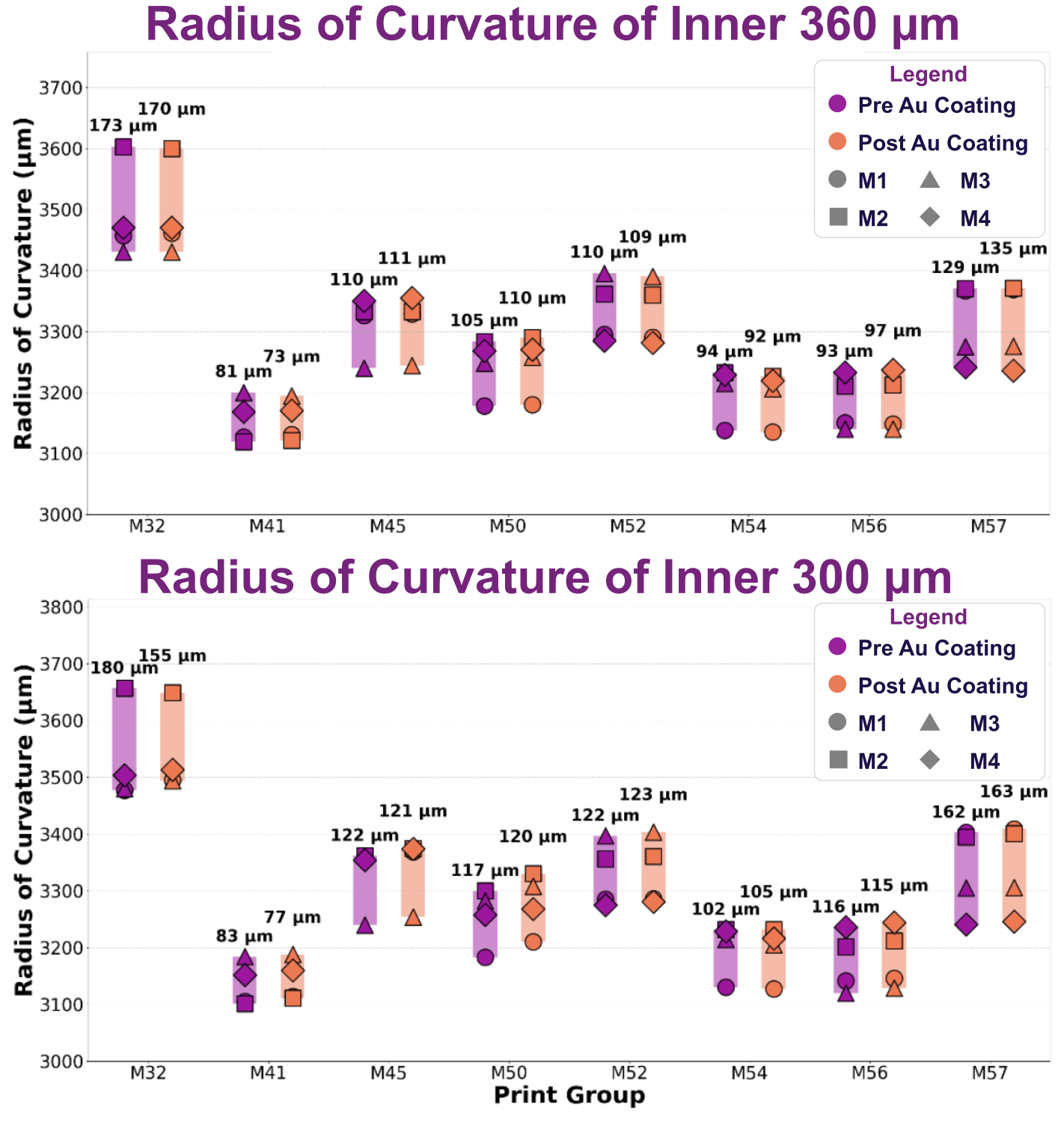}
     \caption{Measured radius-of-curvature of the final mirror arrays pre- and post- Au deposition. Above each array group is the total range of the array's ROC - in bold black font. All arrays met the criteria of falling under the $\pm$ 100 $\mu m$ ROC tolerance. For the inner 360 $\mu m$, the median change is -0.5 $\mu m$. Standard deviation for a mirror is $\pm$4.74 $\mu m$ and for the population is $\pm$4.43 $\mu m$. For the inner 300 $\mu m$ the median change is -3.13 $\mu m$, and the median is -0.5$\mu m$. Sample standard deviation is $\pm$9.17 $\mu m$, and the population standard deviation is $\pm$8.58 $\mu m$. The change experienced by both inner diameter regions is nearly negligible, but requires consideration if an array's ROC range is nearing the $\pm$100 $\mu m$ tolerance. Print M41 and on wards we changed to a new tube of IP-S.}
     \label{fig:Pre and Post Au Coating Final Arrays}
 \end{figure}

~~~~TA-MOONS will observe in the wavelengths 360 nm to 2.5 $\mu m$ and its calibration will be performed in the I-band. Micro-MOONS needs to reflect as much light as possible, and gold has extraordinary reflectivity past 650 nm, far into the infrared. Because of this  gold is chosen as the final metal reflective coating of the mirrors. Due to the poor adhesive properties of gold, an underlayer of titanium is used to help adhere the gold to the mirrors.
The micro-MOONS arrays were coated using a Temescal FC2000 evaporator, E-beam deposition, and a rotating, multi plate fixture to hold the samples within the machine.  The samples were loaded onto the same plate of the rotating fixture, and none of the samples were placed directly in the center of the plate. This is because previous depositions showed that samples set in the center accumulated particles during the evaporation process. 
For the Ti deposition, the set thickness was 0.050 k$\AA$,  at a rate of 2$\AA$ per second, and deposition occurred with the following settings: ion gauge pressure was 1.5E-5 Torr, chamber temperature of 17$^\circ C$, and a mechanical pump pressure of 1.6E-2 Torr. The resulting thickness deposited was 0.053 k$\AA$. 
For the Au deposition, the set thickness was 0.950 k$\AA$, at a rate of 2$\AA$ per second, and deposition occurred with the following settings: ion gauge pressure was 9.7E-6 Torr, chamber temperature of 22$^\circ C$, and a mechanical pump pressure of 1.6E-2 Torr. The resulting thickness deposited was 0.951 k$\AA$. \autoref{fig:Pre and Post Au Coating Final Arrays} shows a graph of the ROC of micro-MOONS arrays before and after deposition.

\section{ CHALLENGES, TESTS \& RESOLUTIONS}

\subsection{Silicon Substrate}
\begin{figure}
    \centering
    \includegraphics[width=1\linewidth]{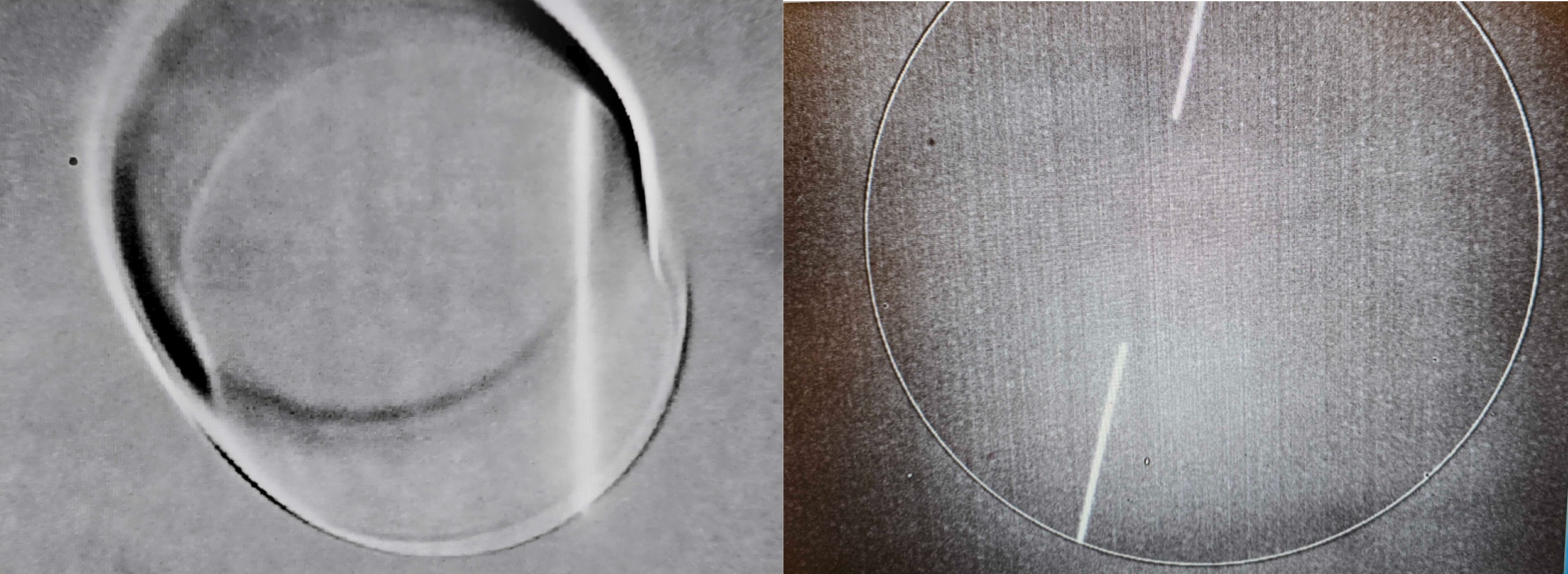}
    \caption{The left image depicts a floating print that failed to adhere to the surface of the substrate. So as it prints, the shape is smeared. The right image is that of an adhered print. Both images are of pulled from the AxioVision Live-View camera software on the NanoScribe computer system.}
    \label{fig:Silicon Drift Print}
\end{figure}

~~~~Initially, when choosing a substrate, indium tin oxide (ITO)-coated glass was considered alongside silicon as both are compatible with the IP-S photoresist. ITO came with a higher cost and more difficulty dicing, while silicon was much more cost-effective and easier to dice. However, silicon came with print difficulties. Known complications with printing on silicon in a NanoScribe stem from its high reflectivity, such as poor adhesion, trouble focusing, potential over polymerization, or uneven voxel distortion. Silicon was preferred due to the ease of dicing the larger substrate to a size compatible with the TA-MOONS pick-up arm, so we underwent problem-solving on how to effectively print onto the material. 
As it is a highly reflective substance, a known issue when printing on silicon is reflecting the incident beam. As this was our main concern, we consulted a NanoScribe proprietary help guide which lays out procedures that can help ease the printing process, such as the addition of a base, and plasma etching to promote adhesion, both of which were already in effect.
Other options, that were more specific for the printing process itself, are optimizing the laser power by means of scaling down the print power from 1 to 0.8, scaling the dosage to 80$\%$. The purpose of scaling down the power is to prevent over polymerization of a region. 
This did not fix our problem as reflection of the incident beam was not our main concern, rather it was focusing onto the substrate's surface. We were printing mirrors in the IP-S, but they were not on the substrate surface (see \autoref{fig:Silicon Drift Print}), instead the mirrors were being printed at the air-photoresist interface (interface is focusing point), the first focusing point. We needed the objective to focus on the photoresist-substrate interface, which is the second focusing point. Within NanoScribe, a terminal override command was found where a user can tell the interface directly to bypass a first focusing point and proceed to a second focusing point, in our case, closer to the substrate's surface. The override is: \texttt{InterfaceMin X.Y}, where X is a measure for the interface signal amplitude, and Y is inversely proportional to the exposure time of the auto focus camera. To determine the correct values for X and Y, we manually found the interface using the rotary control dial, carefully moving the objective z-drive upwards until the interface finder graph showed the expected signal. Once we reached that point, we selected the Find Interface button and used the values reported in the dialog box. We found \texttt{InterfaceMin 1000.090} was consistently effective at locating the correct interface in our system, but recommend that anyone who uses this override find their own X and Y values to avoid any damage to the objective.

\subsection{ROC Deviations}
\begin{figure}
    \centering
    \includegraphics[width=0.80\linewidth]{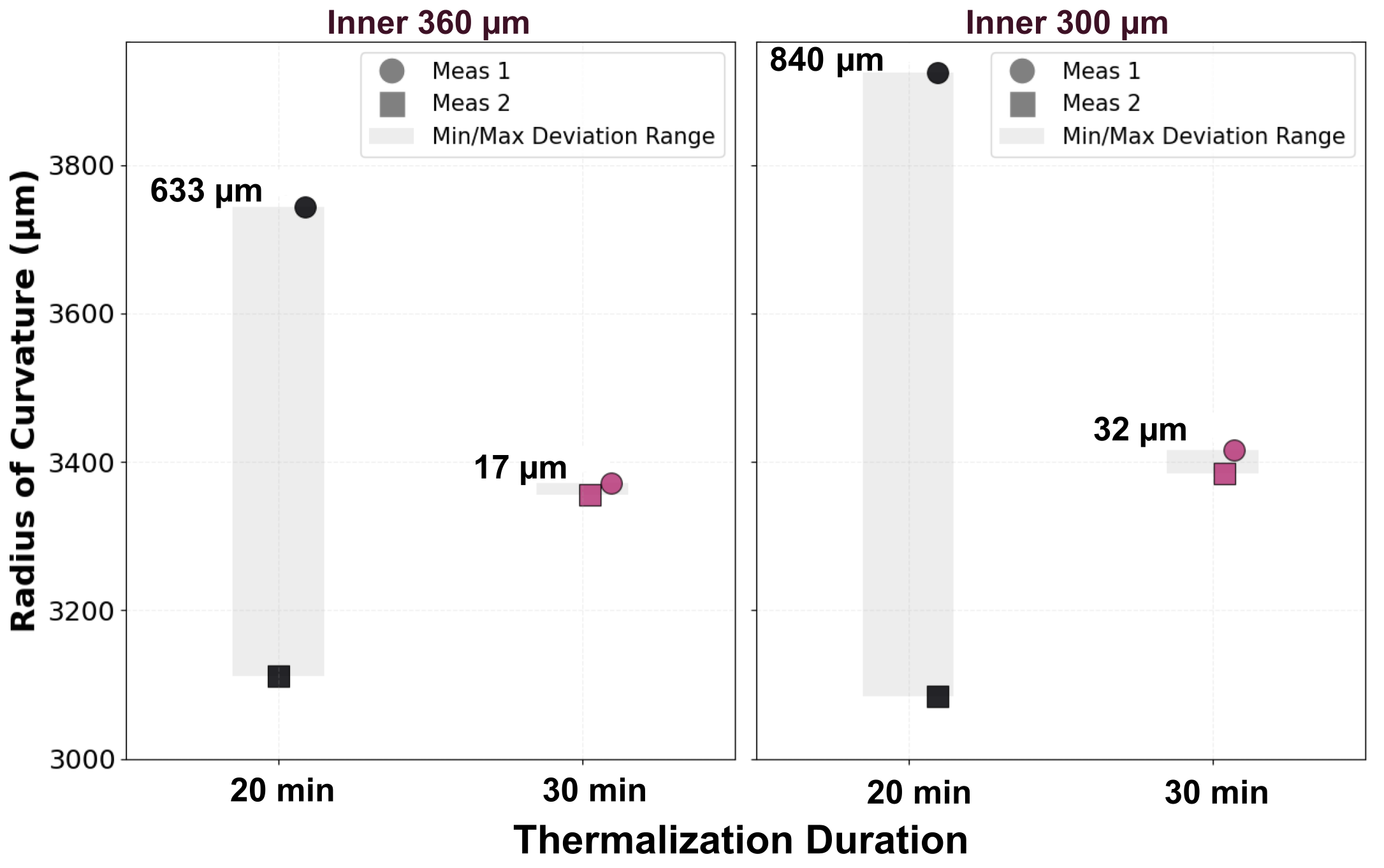}
    \caption{This image shows the stark difference between two mirrors printed within the same 1 by 2 array. There is a 20 minute and a 30 minute thermalization period. One mirror takes about 14 minutes to print, and during that 14 minutes, the 20 minute thermalization print's resin continues to change temperature, leading to a larger difference in the two mirrors' ROC. The 30 minute thermalization period showed consistent results as the difference of ROC was well below the 200 $\mu m$  tolerance.}
    \label{fig:Thermalization Variation}
\end{figure}

~~~~Within arrays that should contain nearly identical prints, the ROC would sometimes vary significantly. Some variations were as great as 800 $\mu m$. As we are seeing limited, spot quality requirements are not imposing rigorous standards, but variation greater than 200 $\mu m$ was not acceptable for the TA-MOONS instrument, based on a Zemax simulation of a spherical mirror. .
To minimize deviation, we first decreased the piezo settling time from 500 ms to 250 ms to preserve the overall shape of the mirror. The change was effective at preserving a more uniform ROC and general shape within the arrays. During our project, we began to properly store the resin in a refrigerated environment, which required an addition to the pre-processing. We began thermalization of the photoresist before printing because the refrigerated storage method would cause mirrors within an array to be printed when the resin is at different temperatures. To determine the best thermalization time, we conducted 2 prints, one set to thermalize for 20 minutes and one set for 30 minutes (see \autoref{fig:Thermalization Variation}{} ). Each print was exactly the same, save for the thermalization period. The best results came from the 30 minute thermalization period, with a difference of ROC between the mirrors being 17 $\mu m$ for the inner 360 $\mu m$ diameter, and a 32 $\mu m$  difference of ROC for the inner 300 $\mu m$ diameter. The addition of a thermalization period allows the photoresist to reach a temperature that will not change during the print duration. The 20 minute print did not provide enough time to reach a stable temperature as the range of ROC within one print was 840 $\mu m$ for the inner 300 $\mu m$. After this test,  we added a 30 minute thermalization period to all prints. 

\subsection{Bubbles}
~~~~Initially, bubbles were not a problem upon opening the IP-S photoresist tube, but over time, they accumulated in the photoresist, becoming an obstacle. We introduced a degassing step to the pre-processing procedure of the print to minimize the impact of bubbles on the prints as seen in \autoref{fig:FESEM Imgages Degassing}. We tested three different degassing periods (10 min, 20 min, and 30 min), and included a 'burp' - a brief release of the vacuum pressure - every 5 minutes in all tests. Results were that the longest degassing time tested, the 30 minute period, yielded the best results. The addition of a de-gassing period removed stray micro-bubbles that lingered near the substrate and instead brought them to the surface of the resin drop. While it did not directly solve the occurrence of all bubbles, it reduced their size upon occurrence (see \autoref{fig:Thermalization Variation}). The primary cause of bubbles is unclear, so to mitigate the problem, the resist was stored upright in a refrigerated environment, which causes some bubbles within the resist to accumulate at the top of the tube instead of towards the dispense nozzle.

\subsection{Aluminum Coated Test Set}
~~~Before making the final arrays that would be coated with gold and sent to the TA-MOONS instrument group, we fabricated 5 test sets and coated them with 100\,nm of aluminum. For additional measurements of the ROC for the micro-MOONS arrays, we used a point source microscope (PSM), a high-precision optical alignment tool that utilizes confocal bright-field imaging to locate the focal points and center of curvature of optical devices \cite{park_optical_2005}. The PSM works by focusing a diffraction-limited point source of light onto a test surface by ways of a microscope objective to measure the point spread function PSF. The light is reflected back through the same objective, imaged onto an internal camera, and the system software tracks the point with high precision. To measure the spot image that will be equivalent to the f/9 beam from DOT telescope, a suitably sized aperture disc was added inside PSM to obtain f/9 beam output. After comparison of the two measuring methods, the PSM measurements tended to be greater then measurements using the optical profilometer, see Table \ref{tab:psm_optical_diff}. 

\begin{table}[h]
\centering
\caption{ The table shows the mean and standard deviation of differences between the PSM and optical profilometer measurements of ROC. While the majority of arrays yielded similar measurements form both devices, the 5th one appears an outlier. It is unclear now what might have cause the large discrepancy in ROC measurement. (PSM $-$ Optical) These measurements are based on the optical profilometer measure of the inner 360 $\mu m$ for ROC.}
\vspace{0.2cm}
\begin{tabular}{cccc}
\hline
Al Coated Array& Mirrors in Array& \textbf{Mean Diff ($\mu m$)} & \textbf{Std Dev ($\mu m$)} \\
\hline
1& 2 & $+69.5$ & $13.4$ \\
2& 4 & $+26.2$ & $32.5$ \\
3& 4 & $-31.0$ & $17.8$ \\
4& 4 & $-0.030.3$ & $51.0$ \\
5& 4 & $+499.2$ & $91.7$ \\
\hline
\end{tabular}
\label{tab:psm_optical_diff}
\end{table}

\begin{figure}
    \centering
    \includegraphics[width=0.98\linewidth]{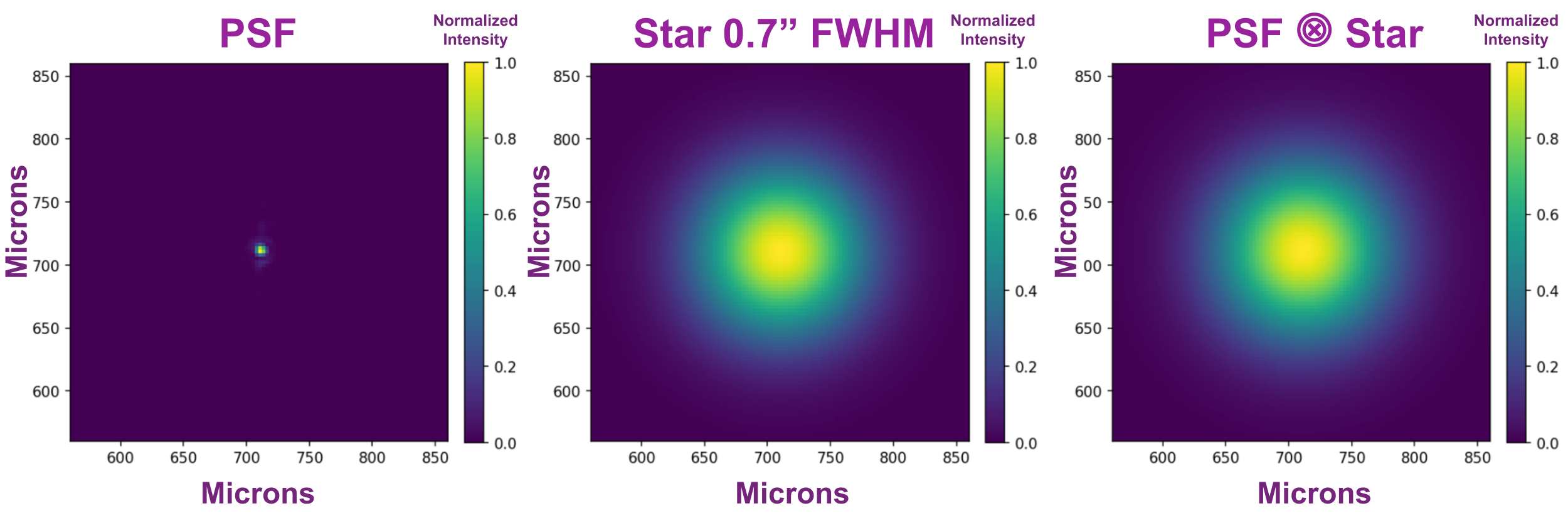}
    \caption{From right to left the images show the measured point spread function (PSF) of a micro-mirror using the PSM at f/9 beam, then the simulated  image of a star at the telescope focal plane (0.7" seeing), then the expected retro reflected star image by the micro-mirror obtained by convolving the measured PSF with the star. The last two figures are similar since the spot diameter of the fabricated mirrors are significantly smaller than the best seeing conditions at DOT. The mirrors therefore do not add any extra aberrations and meets all the requirements of TA-MOONS.}
    \label{fig:Point Source Miscroscope X Star}
\end{figure}

 Additionally, the TA-MOONS team's analysis of the test mirrors concluded that the mirrors are seeing limited, and observation is not constrained by the quality of the mirrors (see \autoref{fig:Point Source Miscroscope X Star}).  
 
\section{CONCLUSION}
\begin{figure}
    \centering
    \includegraphics[width=0.95\linewidth]{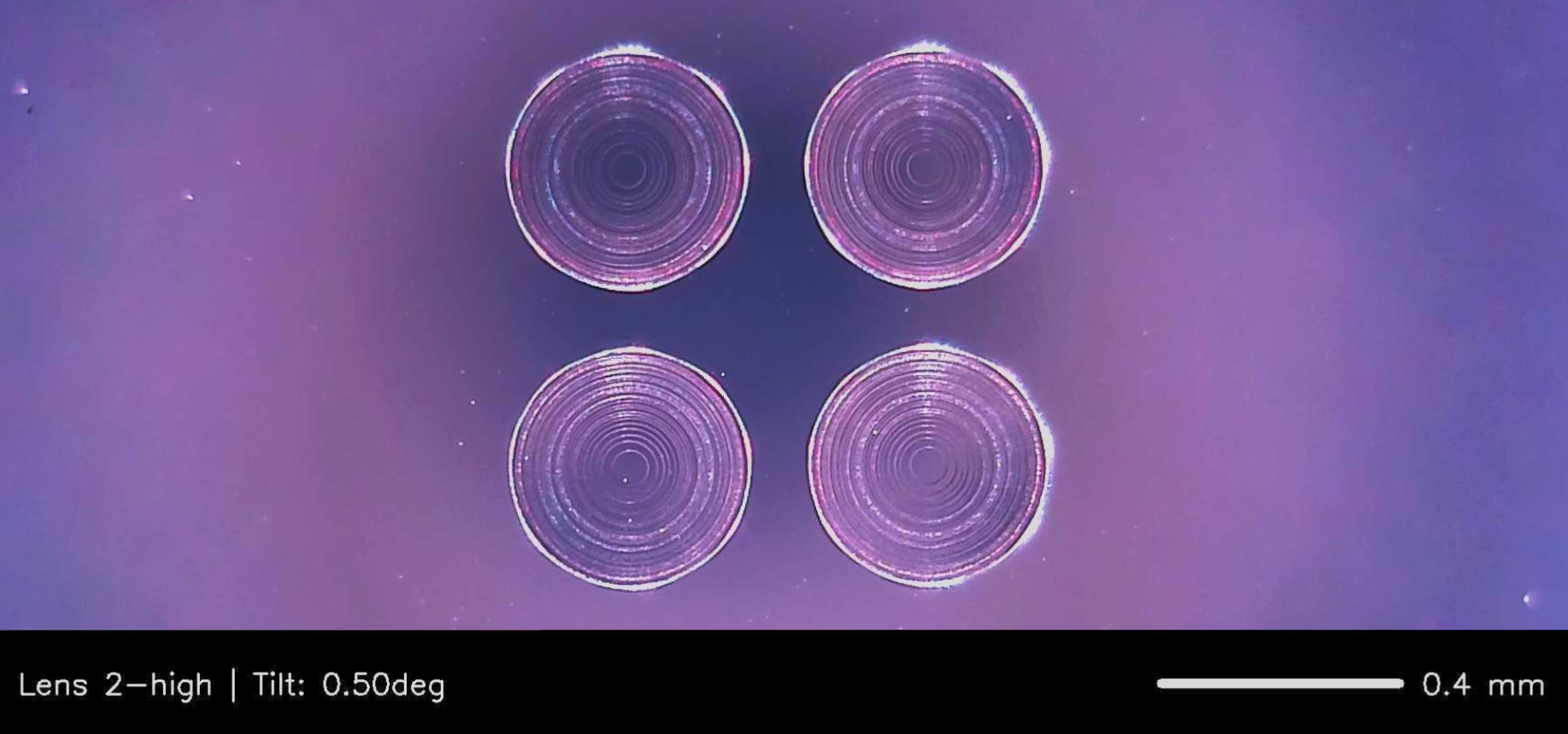}
    \caption{A top-view optical microscopy image of a gold coated micro-MOONS array taken from a Microqubic MRCL700 3D Imager Pro. }
    \label{fig:Au Coated 2x2 Array}
\end{figure}

~~~Two-photon polymerization optics can play an important role in astronomical instrumentation as 2PP allows for custom optic manufacturing on a scale that typical optical manufacturing cannot easily match, or would be very expensive to obtain. The fabrication of the micro-MOONS arrays, with mirrors that are only 400$\mu m$ in diameter, using the process of 2PP was able to meet all the requirements. However, the major challenge that remains is reaching reliable replication of arrays. The appearance of bubbles at the center region of the mirrors, despite taking active measures to remove them, was a constant challenge till the end. The cause of the bubbles is still unknown, or whether a problem stems from the print parameters or the photoresin. Further investigation into replication should be explored. Looking back on prints we conducted after the aluminum test set, and only including the arrays with identical pre-and post processing, 42\% of arrays had at least one mirror affected by a bubble disrupting the surface roughness at the center. Therefore, it is essential to further investigate the replication of 2PP. Addressing this defect rate will be important to prepare 2PP for becoming a dependable manufacturing tool for next-generation astronomical optical instrumentation. Nevertheless, we were able to fabricate eight sets of bubble-free, high-quality mirrors. One such array is seem in \autoref{fig:Au Coated 2x2 Array}. 

\acknowledgments 
 
We acknowledge funding support from the Penn State Materials Research Institute for this program. The Center for Exoplanets and Habitable Worlds is supported by the Pennsylvania State University, the Eberly College of Science, and the Pennsylvania Space Grant Consortium. 

\bibliography{report} 
\bibliographystyle{spiebib} 

\end{document}